\documentclass[twocolumn,pre]{revtex4}

\usepackage{amsmath,amssymb,amsfonts,amsthm}
\usepackage{graphicx}

\newcommand{\bsy}{\bm} 
\renewcommand{\vec}{\mathbf} 
\newcommand{\x}{\vec{x}}
\renewcommand{\u}{\vec{u}}

\newcommand{\ci}{\vec{c}_i} 
\newcommand{\dgC}{\,\mbox{$^\circ$C}} 
\newcommand{\Rey}{Re}

\newcommand{\rt}[1]{{#1}}

\begin{document}

\title{Sedimentation analysis of small ice crystals by Lattice Boltzmann Method}

\author{Juan P. \surname{Giovacchini}$^{1,2}$}
\email{giovacchini@famaf.unc.edu.ar}

\affiliation{$^1$Departamento de Mec\'anica Aeron\'autica, Instituto
Universitario Aeron\'autico, C\'ordoba, Argentina.\\
$^2$Instituto de F\'{\i}sica Enrique Gaviola (CONICET), C\'ordoba,
Argentina.}

\begin{abstract}

Lattice Boltzmann Method (LBM) is used to simulate and analyze the sedimentation
of small ($15-80 \,\mu m$) columnar ice particles in the atmosphere.
We are specially interested in evaluating the terminal falling velocity \rt{of}
columnar ice crystals with hexagonal cross section.
\rt{The main objective is to apply the LBM to solve ice crystal sedimentation
problems. This numerical method is evaluated as a powerful numerical tool to
solve ice crystal sedimentation problems in a variety of sizes.}
LBM results are presented in comparison with laboratory experimental results and
theoretical \rt{approaches} well known in the literature. The numerical results
show good agreement with experimental and theoretical results for the analyzed
geometrical configurations.

\end{abstract}


\maketitle

\section{Introduction}

Ice crystals with a variety of shape, size and mass are present in clouds. The
properties of these crystals are markedly dependent on the temperature and
other properties of the atmosphere \citep{Ryan:1976,Heymsfield:2000}.
A classification of ice crystals with a description of crystal shapes,
size and mass can be found in the works of
\citet{Magono:1966,Ryan:1976,Heymsfield:2000,Bailey:2009,Lindqvist:2012,Um:2009}.

Certain atmospheric and cloud behaviors are characterized by parameters
related to the ice particle dynamics for different shapes and sizes
\citep{Eidhammer:2014}.
A precise estimation of ice \rt{crystal terminal} velocity is required to
quantitatively determine their evolution in the atmosphere.
The knowledge of the \rt{fall} velocity is necessary for the simulation of ice
water \rt{path} and for the determination of cloud boundaries
\citep{Khvorostyanov:2002}.
Also\rt{,} it is used for the study of microphysical process in clouds and for
climate modeling \citep{Kajikawa:1973,Khvorostyanov:2002}.

\rt{It is desirable to obtain accurate and more detailed measurements of
relationships between the terminal velocity, masses, and dimensions for a large
spectrum of ice crystal \rt{shapes}.
A precise determination of these relationships allow us to obtain more reliable
\rt{terminal} velocity parameterizations of cloud particles.
These parameterizations are essential to have an accurate simulation of
\rt{clouds} in general circulation models (GCMs) of precipitation amount, cloud
dissipation and cloud optical properties
\citep{Khvorostyanov:2002,Heymsfield:2000}.}

Although there have been many proposals in \rt{the} literature, the ice crystal
sedimentation in the atmosphere has not been completely characterized
\citep{Heymsfield:2000,Westbrook:2008,Heymsfield:2010}.
There are analytical solutions that precisely determine the \rt{terminal
falling velocity} for \rt{spherical} particles.
Due to the large variety of shapes, sizes and masses of ice crystals, and the
range of Reynolds numbers involved in these problems, there is no precise
analytical estimation to predict the \rt{terminal} velocity for shapes other
than spheres.
Many works in literature
\citep{Bohm:1989,Bohm:1992,Mitchell:1996,Mitchell:2005,Heymsfield:2000,
Khvorostyanov:2002,Khvorostyanov:2005,Westbrook:2008,Heymsfield:2010}
provide schemes to parameterize the ice crystal masses, shapes and size to
predict the \rt{terminal} velocity.
Ice particle terminal velocities are often calculated theoretically or
experimentally by determining a relationship between the Reynolds number
($\Rey$), and the Best (or Davis) number, ($X$)
\citep{Jayaweera:1969,Jayaweera:1972,Bohm:1989,Mitchell:1996}.

There are a number of experimental works in which the most important variables
are measured.
Terminal velocity, mass and size have been measured for various ice particle
types.
These data-sets are obtained from laboratory measurement and observations of
real ice particles falling through the atmosphere.
Some well known experimental data-sets can be found in
\cite{Jayaweera:1969,Jayaweera:1972,Kajikawa:1973,Locatelli:1974,Michaeli:1977,
Burgesser:2016}.

The proposals by \citet{Bohm:1989,Bohm:1992,Mitchell:1996,Mitchell:2005,
Khvorostyanov:2002,Khvorostyanov:2005} have shown quite good approximations to
the \rt{terminal velocity} of ice particles for $\Rey \gg 1$.
These proposals proved to be in good agreement with experimental data for a
variety of particle types.
However \citet{Westbrook:2008b} showed that for viscous flow regimes
($\Rey \ll 1$) these formulations overestimate the crystal \rt{terminal}
velocity.
\citet{Westbrook:2008}, using the approximation of \citet{Hubbard:1993} with
results from \cite{Westbrook:2008b}, \rt{gives} an estimate for the
sedimentation rate of small ice crystals whose maximum dimension is smaller than
$100 \mathrm{\mu m}$.
This estimate for columnar ice crystals is in agreement with most experimental
data (within 20\%).

A complete review of the main theoretical \rt{approximations} that have been
proposed and many experimental results can be found in
\citep{Mitchell:1996,Mitchell:2005,Heymsfield:2000,Khvorostyanov:2002,
Khvorostyanov:2005,Heymsfield:2010};
also a lots of relevant references are presented in these works.

The sedimentation of an ice crystal in the atmosphere is a fluid mechanical
problem that can be modeled as a rigid body moving immersed in a fluid flow.
This rigid body moves under the action of its own weight, buoyancy force and
interaction with \rt{other} crystals and with the fluid that surrounds it.

Given a characterization for the shape, size and mass density of the crystal,
together with the \rt{atmospheric conditions}, it is possible to completely
determine \rt{its} dynamical behavior by using some adequate computational
fluid \rt{dynamics} (CFD) method.
\rt{An accurate numerical method allows us to compute the
terminal velocities for sizes, shapes and masses for which experimental data are
not available. Also, for given shapes the sensitivity of the problem related to
sizes and masses can be studied numerically.}

\rt{To the knowledge of the author, there are no numerical results studying the
sedimentation of ice crystals for the range of lengths
($l=15-80 \mathrm{\mu m}$) we study in this paper.
However, there are some approaches in the literature to numerically solve ice
crystal sedimentation problems in the atmosphere
\citep{Cheng:2013,Cheng:2015,Hashino:2016}.
\citet{Hashino:2016} study the sedimentation of columnar crystals with
$l > 600 \mu m$ using a commercial software (ANSYS Fluent) that
uses finite volume methods applied to the Navier-Stokes equation.}


The \rt{Lattice Boltzmann method} (LBM) is a CFD method that proved to be
successful to treat multiple problems involving both compressible and
quasi-incompressible flows on simple and complex geometrical settings.
In particular, the LBM provide a simple way for treating accurately the flow
surrounding an immersed body, in arbitrary movement, with no regular geometry.
For a complete modern review of this topic see \cite{Aidun:2010}.
The behavior of particles in sedimentation have been analyzed using LBM in a
variety of problems \citep{Ladd:1994,Ladd:1994b,Aidun:1995,Aidun:1998,Xia:2009}.

In this paper we use LBM to study the dynamical behavior of columnar ice
crystals.
\rt{The ice crystal terminal velocity is obtained numerically for a range of
sizes, characteristic lengths in range $l=15-80 \mathrm{\mu m}$.}
The LBM results for the fluid mechanical problems are obtained in a
pure viscous regime ($\Rey \ll 1$). This is the flow regime of the smallest
particles falling in a cloud.
The accuracy in the LBM to treat this problem is evaluated by comparison with
some well known experimental data in literature
\citep{Jayaweera:1972,Kajikawa:1973,Michaeli:1977,Burgesser:2016}, and with
theoretical proposals from
\citep{Mitchell:1996,Mitchell:2005,Khvorostyanov:2002,Khvorostyanov:2005,
Westbrook:2008}.

The paper is organized as follows:
In section \ref{sec:LBM} we present the basic equations of the LBM, introduce
notation and some details about the boundary conditions methods, force
evaluation, and grid refinement techniques.
In section \ref{sec:IceParticleSedimentation} the sedimentation of ice crystals
in the atmosphere is solved using LBM. Numerical results for \emph{columnar ice
crystals} are \rt{shown} in sections \ref{subsec:ColumnarIceParticles}.
In section \ref{sec:ConclusionAndDiscussion} conclusions and discussions are
presented.
In the appendix \ref{AppA:GridConvergence} we check the convergence of the
method with respect to the grid size.
In the appendix \ref{AppB:LBMresults} we present tables with the obtained LBM
results.

\section{The lattice Boltzmann method}\label{sec:LBM}

In this section we present the basic equations of the LBM, introduce notation
and the main concepts we use along the paper.

In addition to the lattice Boltzmann equation that govern the physics of the
bulk fluid; one needs to prescribe a method to apply boundary conditions, to
evaluate the fluid force on a body and to implement grid refinement where
necessary. In the next sections we briefly review these topics.

\subsection{Lattice Boltzmann equation}

The numerical results in this paper are obtained by solving the lattice
Boltzmann equation (\emph{LBE})
\citep{dHumieres:1992,He:1997,Succi:2001,Wolf-Gladrow:2000}, a particular
phase-space and temporal discretization of the Boltzmann equation (\emph{BE})
\citep{Harris:2004,Sone:2007}.

The BE governs the time evolution of the single-particle distribution function
$f(\x,\bsy{\xi},t),$ where $\x$ and $\bsy{\xi}$ are the position and velocity of
the particle in phase space.
The LBE is a discretized version of the BE, where $\x$ takes values on a uniform
grid (the lattice), and $\bsy{\xi}$ is not only discretized, but also restricted
to a finite number $Q$ (the number of discrete velocities in the model) of
values \citep{He:1997}.
In an isothermal situation and in the absence of external forces, like gravity,
the LBE can be written as:

\begin{multline}\label{eq:LBE}
  f_i(\x_A+\ci\delta t, t+\delta t)  = f_i(\x_A, t) + \\
  \sum_{j=0}^{Q-1}\bsy{\Omega}_{i,j} \Bigl( f_j(\x_A, t) - f_j^{eq}(\x_A, t) \Bigr)  \quad , \quad  \\ i=0,1,\dots,Q-1.
\end{multline}
Here $f_i=f_i(\x_A, t)$ is the $i$-th component of the discretized distribution
function $\bsy{f}(\x_A, t)$ at the lattice site $\x_A,$ time $t,$ and discrete
velocity $\ci$.
The coordinates of a lattice node are denoted by $\x_A$, where the integer multi
index $A=(j,k,l)$ (or, $A=(j,k)$ in the two-dimensional case) denotes a
particular site in the lattice.
The function $\bsy{f}^{eq}(\x_A, t)$ is an equilibrium distribution function and
$\bsy{\Omega}$ is a linearized collision operator.
In our simulations we use a multiple relaxation time model (\emph{MRT})
\citep{dHumieres:1992,dHumieres:2002}.
The setting of the model used, such as relaxation parameters, equilibrium
expressions and others, are those proposed in \cite{dHumieres:2002}.

The macroscopic quantities such as the fluid mass density $\rho(\x_A,t),$ and
velocity $\u(\x_A,t)$, are obtained as usual in lattice Boltzmann theory
\citep{He:1997,Wolf-Gladrow:2000}.

We refer to the lattice Boltzmann models with the standard notation $DmQn$,
where $m$ is the number of space dimensions of the problem, and $n$ is the
number of discrete velocities.
To obtain the results we use the D3Q15 velocity model.

\subsection{Boundary conditions}\label{subsec:LBM_BC}

The problems we are interested in are those in which rigid bodies move inside an
unbounded fluid domain.  Because of the impossibility to model an infinite fluid
domain, we have to restrict the problem to a finite computational fluid domain.
The size of the computational fluid domain has to be a compromise between
minimizing the computational work---the smaller the size the better, and
minimizing the undesirable effect of the boundary conditions---the larger the
domain the better.

The computational fluid domain is a block of fluid bounded by regular borders.
The rigid bodies that move inside the domain are described by geometries as
required.

The flow in the interior of the domain is computed by solving the LBE. Close to
the boundaries a special treatment is used so that the flow obeys the physical
boundary conditions. In the present work, we use both Dirichlet and outflow
open-boundary conditions (convective boundary conditions or Sommerfeld like
conditions).
The correct imposition of the boundary conditions on arbitrary boundary
geometries, like the boundary of rigid bodies, has been one of the main
issues in LBM development.

Dirichlet velocity boundary conditions on boundaries of arbitrary shape are
imposed by the method proposed in \citet{Bouzidi:2001b}.

We use outflow open-boundary conditions to represent a long or quasi-infinite
physical domain by a finite computational domain.
These type of conditions have been extensively applied in computational fluid
mechanics.

There are different approaches in the literature to treat the outflow
open-boundary conditions in the LBM context. We can divide these
approaches in at least two categories, the ones based on mesoscopic variables
\citep{Yu:2005,Chikatamarla:2006} and the \rt{others} based on macroscopic
variables \citep{Aidun:1998,Yang:2007,Junk:2008,Yang:2013}.
\rt{The last group of references are} generally extensions of boundary
conditions extensively applied in classical methods (Finite Difference
(\emph{FD}), Finite Volume (\emph{FV}) and Finite Element (\emph{FE}) methods)
of computational fluid mechanics (\emph{CFD}) to solve the Navier-Stokes
(\emph{NS}) equations.

We are mainly interested in non-stationary quasi-incompressible problems.  In
the LBM context the convective boundary condition (\emph{CBC}) proposed in
\citet{Yang:2013} to treat outflow open-boundaries has shown acceptable results
in these kind of problems.
The Neumann boundary conditions (\emph{NBC}) were also tested in LBM
\citep{Aidun:1998,Yang:2007,Junk:2008,Yang:2013}. The results presented in
\citet{Yang:2007,Yang:2013} show that CBC \rt{is} a better option than NBC in
non-stationary problems. These works show that NBC introduce undesirable
perturbations in the fluid domain, specially in non-stationary problems.

In our numerical tests we use CBC method as proposed by \citet{Yang:2013} to
treat the outflow open-boundaries.

\subsection{Forces evaluation}\label{subsec:LBM_ForceAndTorque}

It is of crucial importance, in many applications that involve moving bodies
surrounded by a fluid flow, to have a good method or algorithm to compute the
flow force and torque acting on the bodies. By good we mean a method that is
simple to apply, that is accurate and fast, so as not to spoil the efficiency of
the flow computing method.
The accuracy in the determination of the force and torque acting on a moving
body directly affects the body's movement.
For a review of LBM methods that involve flow force evaluation on suspended
particles we refer to Section 6 of \citet{Aidun:2010} and references therein.

The classical way to compute forces, and so torque, on submerged bodies is via
the computation and integration of the stress tensor on the surface of the body.
In LBM the stress tensor is a local variable, its computation and extrapolation
from the lattice to the surface is computationally expensive, which ruins the
efficiency of the LBM. However, this method is widely used in LBM
\citep{Inamuro:2000, Xia:2009, Li:2004}.

\rt{A} standard method to evaluate forces on submerged bodies in LBM is the
\emph{momentum exchange} (\emph{ME}), introduced firstly by
\citet{Ladd:1994, Ladd:1994b} in LBM applications.
The ME algorithm is specifically designed and adapted to LBM; it is therefore
more efficient than stress integration from the computational point of view.

Some improvements to \rt{the Ladd method} have been introduced in
\cite{Aidun:1995,Aidun:1998,Mei:2002}, and different approaches to improve the
methods in problems with moving bodies were made
\citep{Wen:2012,Wen:2014,Giovacchini:2015}. In this work, force and torque are
evaluated by using the methods presented in \citet{Giovacchini:2015}.

The motion of each body is determined by solving the Newton's equations of
motion.
The forces acting over bodies are given by the fluid flow forces, weight and
buoyancy forces.
To integrate in time we use Euler Forward numerical scheme, which is first order
accurate as the LBM method itself.

\subsection{Grid refinement method}\label{subsec:GridRefinementMethod}

Many problems in fluid mechanics are such that \rt{large} gradients of the fluid
variables appear only in regions which are small compared \rt{to} the whole
computational domain. To resolve well the space variations of the fluid
variables\rt{,} one needs a grid size which is small enough.

In LBM a simple lattice is a Cartesian grid of equispaced nodes. The distance
between two nearest neighbor nodes, the grid size, is $\delta x.$ For a
real problem, the computational domain is covered by an
arrangement of grids. This arrangement can be as simple as a unique lattice---or
block grid---with a single size $\delta x$, or a complex arrangement of grids
with different grid sizes.

In a problem with more or less uniform space variations \rt{throughout}, a
single block grid that covers the whole computational domain may be suitable. In
a problem where high space variations occur in a small region, a small grid size
needs to be used in that region. But using this small grid size on the whole
computational domain would be a waste of computational effort. The right thing
to do is to use an arrangement of grids with different grid sizes. The methods
to integrate various grid blocks with different grid sizes into a single
computational domain are known as grid refinement methods.

In LBM there are at least two grid refinement methods: \emph{multi-grid method}
(MG) \citep{Filippova:1998,Filippova:1998b} and \emph{multi-domain method} (MD)
(or multi-block) \citep{Yu:2002,Lagrava:2012}.
In the MG method, a grid block with small grid size is always superimposed to a
grid block with larger grid size. Several layers of grids can be superimposed
in this way.
In MD method the grids with different grid sizes overlap just in a selected set
of lattice nodes. This overlapping occurs only on a small region with two
adjacent grid blocks of different grid sizes (see \citep{Yu:2002,Lagrava:2012}).

In this work we use MD methods. We select this method because it has better
numerical performance and lesser memory requirement than MG method. A
disadvantage of the MD method, though, is that its implementation is more
complex than that of MG where some additional grids are used as interface to
interchange data between different levels of grid size.

We use \rt{an} a priori refinement method. This means that we chose the
arrangement of refined grids in the domain before solving the fluid problem.
The region where the refinement is applied is not static.
We implement an algorithm to follow the rigid body, so that the body is
approximately centered in the refinement region at all times.

\section{Ice crystals sedimentation - numerical results}
\label{sec:IceParticleSedimentation}

In this section\rt{,} we study the main problem of this paper, we solve and
analyze the ice crystals sedimentation in the atmosphere.
In particular, we are interested in evaluating the ice crystal \rt{terminal}
velocity by using LBM for columnar ice crystal shape in a size range.

The sedimentation of an ice crystal in the atmosphere is a fluid mechanical
problem \rt{that} we model as follows. The crystal is considered a rigid body
that moves under the action of its own weight, the buoyancy force and
interacting only with the fluid that surrounds it.
A simplifying assumption is adopted: no interactions between rigid bodies is
considered.
We are only interested in isolated rigid bodies in the atmosphere.
This assumption is a good approximation to the movement of ice crystals in a
cloud, since the concentration of ice particles in cirrus typically ranges
between 50 and 500 $\mathrm{liter^{-1}}$, while the maximum ice particle
concentration in cumulonimbus clouds \rt{reaches} 300 $\mathrm{liter^{-1}}$
\citep{Pruppacher:1998}.
\rt{It should be noted that the concentration of ice particles can be higher in
anvil clouds.}

The \rt{results obtained with LBM} are compared with some well known
experimental data in \rt{the} literature
\citep{Jayaweera:1972,Kajikawa:1973,Michaeli:1977,Burgesser:2016}, as much as
with the theoretical proposals from
\citep{Mitchell:1996,Mitchell:2005,Khvorostyanov:2002,Khvorostyanov:2005,
Westbrook:2008}.

\subsection{Columnar ice crystals}
\label{subsec:ColumnarIceParticles}

Columnar ice crystals with quasi-hexagonal cross section and needle ice
crystals, \rt{are typically grown at temperatures in the ranges
$-3$ to $-9 \dgC$ and $-18$ to $-24 \dgC$
\citep{Magono:1966,Ryan:1976,Heymsfield:1984,Westbrook:2008,Bailey:2009}.}

\begin{figure}
\centering
\includegraphics[scale=0.25]
{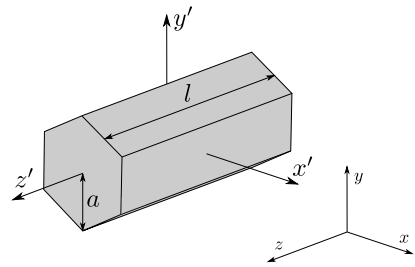}
\caption{Crystal rigid body scheme used to represent the columnar ice crystal
with hexagonal cross section.
$x', y', z'$ are the cartesian coordinates in a frame fixed to the body, while
$x, y, z$ is a fixed inertial coordinate system.}
\label{fig:IceCrystalsRigidBodyScheme}
\end{figure}

In our simulations the ice crystals are modeled \rt{as} columns of hexagonal
cross section (see figure \ref{fig:IceCrystalsRigidBodyScheme}).
The sedimentation is studied in fluid flow regimes with $0.006 < \Rey < 0.4$.
This is approximately the flow regime of the smallest ice particles falling in a
cloud.
In \rt{Figure} \ref{fig:IceCrystalsRigidBodyScheme}, we show \rt{a schematic}
rigid body representing an ice crystals.
We denote with $l$ the ice crystal length and $a$ is the semi-length of its
cross section.
$x', y', z'$ are the cartesian coordinates in a frame fixed to the body.
The rigid body spatial orientation with respect to a fixed coordinate system
$x, y, z$ is defined by the Euler angles $\phi, \theta$ and $\psi$ following the
$z,x,z$ intrinsic rotational order.
We adopt $d=\sqrt{l^2+(2a)^2}$ as a reference length and we use it to evaluate
the Reynolds number.

We perform numerical tests for a variety of aspect ratios
$a_{r}=\frac{l}{2a}\in[1,3]$.
The length of the columnar ice crystals analyzed are in the range
$15\,\mathrm{\mu m}\le l \le 80\,\mathrm{\mu m}$.
The fluid properties are set as those of air at temperature $T=-8\dgC$ with an
atmospheric pressure $P=101325\,$Pa.
We have selected these fluid properties to simulate the atmospheric laboratory
conditions used in \citet{Burgesser:2016}.

\subsection{Results and \rt{discussion}}
\label{subsec:ResAndDisc}

\begin{figure}
\centering
\includegraphics[scale=1.0]{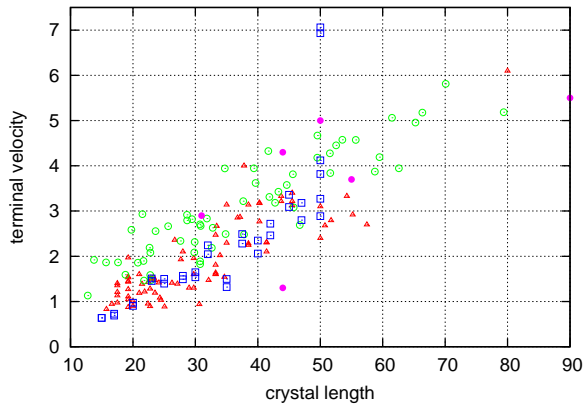}
\caption{\rt{Terminal} velocity $v_{c}$ for columnar ice crystals as a function
of their length $l$. The results correspond to $a_{r}\in[1,2]$.
Velocity is expressed in $\mathrm{cm\,s^{-1}}$, \rt{size} in $\mathrm{\mu m}$.
The LBM results \rt{with $\rho_{ice}=800\mathrm{kg \, m^{-3}}$} are \rt{shown}
in squares ($\Box$).
The experimental results are represented with: triangles ($\triangle$) for data
from \citet{Burgesser:2016}, hollow circles ($\circ$) for \cite{Kajikawa:1973}
data, and with filled circles ($\bullet$) for \citet{Jayaweera:1972} data.
The LBM numerical results are presented in appendix \ref{AppB:LBMresults}.}
\label{fig:ResColumnarExperimental_ar=1_2}
\end{figure}

\begin{figure}
\centering
\includegraphics[scale=1.0]{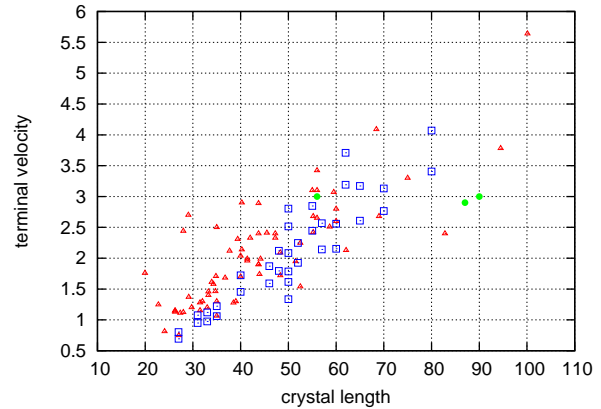}
\caption{\rt{Terminal} velocity $v_{c}$ for columnar ice crystals as a function
of the length $l$. The results correspond to $a_{r}\in[2,3]$.
Velocity is expressed in $\mathrm{cm \, s^{-1}}$, \rt{size} in $\mathrm{\mu m}$.
The LBM results \rt{with $\rho_{ice}=800\mathrm{kg \, m^{-3}}$} are \rt{shown}
in squares ($\Box$).
The experimental results are represented with: triangles ($\triangle$) for data
from \citet{Burgesser:2016}, and with filled circles ($\bullet$) for
\citet{Jayaweera:1972} data.
The LBM numerical results are presented in appendix \ref{AppB:LBMresults}.}
\label{fig:ResColumnarExperimental_ar=2_3}
\end{figure}

In \rt{Figures} \ref{fig:ResColumnarExperimental_ar=1_2} and
\ref{fig:ResColumnarExperimental_ar=2_3} we show the \rt{terminal} velocity we
obtained using LBM in comparison with the laboratory experimental results
presented by \citet{Jayaweera:1972,Kajikawa:1973,Burgesser:2016} as a function of
crystals length.
In \rt{Figure} \ref{fig:ResColumnarExperimental_ar=1_2}, LBM and experimental
results are presented for aspect ratios between 1 and 2; while in \rt{Figure}
\ref{fig:ResColumnarExperimental_ar=2_3} comparative results for aspect ratios
between 2 and 3 are \rt{shown}.
In the appendix \ref{AppB:LBMresults} we include tables with the LBM results.

The \rt{LBM results presented} for each geometrical configuration were obtained
for hexagonal columns with two orientations, horizontal and vertical
($(\phi,\theta,\psi)=(0,0,0)$ and $(0,\frac{\pi}{2},0)$ respectively).
These orientation are expected to produce the lower and upper limits for the
\rt{terminal} velocities.
As opposed to performing simulations for many different orientations, this
strategy allows us to reduce the computational cost. This is particularly true
for problems that do not show a preferred sedimenting orientation.

The ice density for columns are set in the range reported by \citet{Ryan:1976}
for $T=-8 \dgC$. It is also possible to obtain the ice crystal mass from
relationships like those shape based proposed by \citet{Mitchell:1996}.
\rt{\citet{Heymsfield:2000} propose $\rho_{ice}=810\mathrm{kg\, m^{-3}}$ for
columns.
We use two definite values of ice density, $\rho_{ice}=800\mathrm{kg \, m^{-3}}$
for almost all tests, and $\rho_{ice}=400\mathrm{kg \, m^{-3}}$ in some
particular cases.
We use the latter value (approximately the mass density of hollow columns) to
test the dependence of a normalized \rt{terminal} velocity on the mass density.}

As can be observed in Figures \ref{fig:ResColumnarExperimental_ar=1_2} and
\ref{fig:ResColumnarExperimental_ar=2_3}, the LBM results \rt{with
$\rho_{ice}=800\mathrm{kg \, m^{-3}}$} are in accordance with laboratory
measurements.
The results from \cite{Jayaweera:1972,Kajikawa:1973,Burgesser:2016} present
some dispersion as expected for a set of experimental data.
For the length range and aspect ratio analyzed, all the numerical results are
included in the data dispersion.
\rt{In Figure \ref{fig:ResColumnarExperimental_ar=1_2}, two test values for
$l=50 \mathrm{\mu m}$ with $a_{r}=1.0$ in horizontal and vertical sedimentation
need special consideration.
This is not a problem of the LBM simulation. The explanation is that the
experimental data in Figure \ref{fig:ResColumnarExperimental_ar=1_2} do not
contain values with such small aspect ratio and length $l=50 \mathrm{\mu m}$.
Ice crystals with such geometrical configuration were observed in nature as
presented by \citet{Um:2015}, but these were not observed in the experimental
data presented in Figure \ref{fig:ResColumnarExperimental_ar=1_2}.}

In laboratory experiments, the ``measured length'' $l'$ of falling crystals is
in fact a projection of the actual length on a vertical plane.
Thus, the presented length can actually be an underestimation of the real
crystal length $l$ (see details in \cite{Burgesser:2016}).
Owing to this, the experimental data in Figures
\ref{fig:ResColumnarExperimental_ar=1_2} and
\ref{fig:ResColumnarExperimental_ar=2_3} \rt{might} present a bias to the left.
\citet{Burgesser:2016} report differences between the mean value of $l$ and $l'$,
this is 25\% for $2a \in [5,15]\mathrm{\mu m}$, 17\% for
$2a \in [15,25]\mathrm{\mu m}$, and 13\% for $2a \in [25,35]\mathrm{\mu m}.$
The mean values of $l$ are larger than those of $l'$ except for
$2a \in [25,35]\mathrm{\mu m}$.

\begin{figure}
\centering
\includegraphics[scale=1.0]{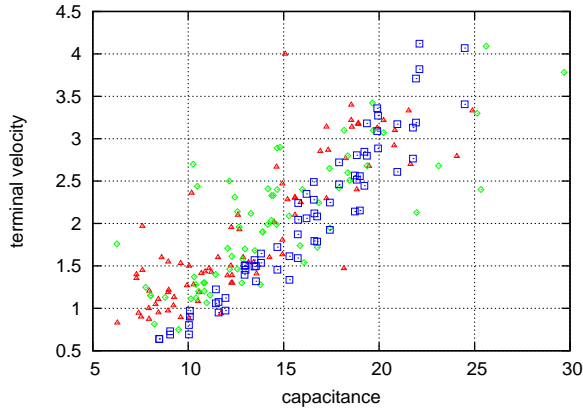}
\caption{\rt{Terminal} velocity $v_{c}$ for columnar ice crystals as a function
of the capacitance $C$.
Velocity is expressed in $\mathrm{cm \,s^{-1}}$, $C$ in $\mathrm{\mu m}$.
The LBM results \rt{with $\rho_{ice}=800\mathrm{kg \, m^{-3}}$} are shown in
squares ($\Box$).
The experimental data from \cite{Burgesser:2016} are represented as:
triangles $\triangle$, for aspect ratios between 1 and 2, and diamonds
$\diamond$ for aspect ratios between 2 and 3.
The LBM numerical results are presented in appendix \ref{AppB:LBMresults}.}
\label{fig:Results_InC}
\end{figure}

In \rt{Figure} \ref{fig:Results_InC} the \rt{terminal} velocities obtained with
LBM are shown in comparison with laboratory measurements presented in
\citep{Burgesser:2016} as a function of crystals capacitance $C$.
This parameter depends on the particle geometry and is obtained in
\cite{Westbrook:2008b} for different geometries.
For hexagonal columns the capacitance is:
\begin{equation}\label{eq:ColumnarCapacitance}
  C = 0.58 a \left( 1 + 0.95 a_{r}^{0.75} \right)
\end{equation}

The numerical results are in the regime which allow us to compare with the
measurements presented in \cite{Burgesser:2016}, where they choose $C$ as the
characteristic length to evaluate the Reynolds number.
The Reynolds number regime of the LBM results is $0.002 < \Rey < 0.15$ if we
take $C$ as the characteristic length.

It is possible to observe from \rt{Figure} \ref{fig:Results_InC} that the
dispersion of the LBM results have a decrease when the capacitance is used as
variable.  The same observation was pointed out in
\citep{Burgesser:2016} for the experimental results.  We can also observe from
Figure \ref{fig:Results_InC} that the results obtained with LBM are not
uniformly distributed within the region containing experimental data.
A bias towards the low part in this region can be seen.
This \rt{may} be explained by the difference between $l$ and $l'$, particularly
for small values of $2a$.
\rt{In Figure \ref{fig:Results_InC} the LBM results for $l=50 \mathrm{\mu m}$
and $a_{r}=1$ are not shown because these extend outside of data dispersion as
explained above.}

\begin{figure}
\centering
\includegraphics[scale=1.0]{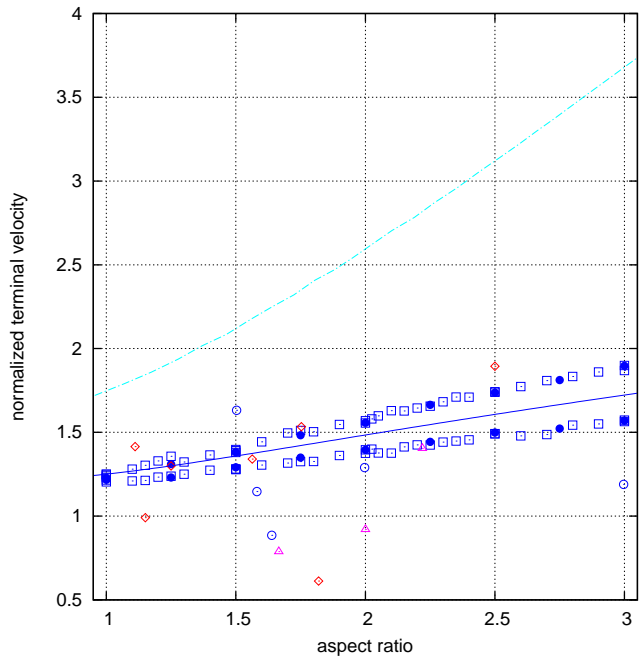}
\caption{Normalized \rt{terminal} velocity $v_{n}$ for hexagonal columns in
function of ice crystals aspect ratio $a_{r}$.
\rt{The LBM results are shown with squares ($\Box$) and filled circles
($\bullet$) for $\rho_{ice}=800\mathrm{kg\,m^{-3}}$ and
$\rho_{ice}=400\mathrm{kg\,m^{-3}}$ respectively. The LBM numerical results are
presented in appendix \ref{AppB:LBMresults}.}
The continuum line shows the normalized terminal velocity proposed in
\cite{Westbrook:2008} for a random orientation ice crystals.
We use triangles $\triangle$ for \cite{Michaeli:1977}, diamonds $\diamond$ for
\cite{Jayaweera:1972}, and circles $\circ$ for \cite{Kajikawa:1973} experimental
results.
The dash-dotted line is the MHKC proposal
\citep{Mitchell:1996,Mitchell:2005,Khvorostyanov:2002,Khvorostyanov:2005}
(see the text).
}
\label{fig:ResColumnarIceParticle}
\end{figure}

In Figure \ref{fig:ResColumnarIceParticle} we show the normalized \rt{terminal
velocity} $v_{n}$ obtained by LBM in comparison with some well known theoretical
and experimental results from the literature.
These results are the same \rt{as we have presented before but in a normalized
way}.
\rt{In addition, we include in \rt{Figure} \ref{fig:ResColumnarIceParticle} the
results obtained for $\rho_{ice}=400\mathrm{kg \, m^{-3}}$.}
The normalized \rt{terminal} velocity is computed as proposed by
\cite{Westbrook:2008}, this is, to obtain $v_{n}$, the crystal \rt{terminal}
velocity $v_{c}$ is divided by a \rt{terminal sedimentation velocity $v_{r}$ of}
an ``equivalent sphere''.
Here, equivalent sphere means a sphere with diameter $d$ and mass $m_{s}$ equal
to the ice crystal mass $m_{ic}$.

\citet{Westbrook:2008}, based \rt{on} results from \cite{Hubbard:1993}, propose
an expression for $v_{c}$ using the Stokes \citep{White:2005} solution for an
sphere in a viscous flow where the sphere radius is replaced by an effective
hydrodynamic radius proportional to the capacitance $C$.
The continuum line in \rt{Figure} \ref{fig:ResColumnarIceParticle} is the
theoretical proposal presented by \citet{Westbrook:2008} to the normalized
\rt{terminal} velocity. This proposal is formulated for columnar ice crystals
with hexagonal cross section in random orientation.
The dash-dotted line in \rt{Figure} \ref{fig:ResColumnarIceParticle}, labeled as
MHKC, \rt{corresponds} to the proposals from
\citet{Mitchell:1996,Mitchell:2005,Khvorostyanov:2002,Khvorostyanov:2005} in
random orientation. MHKC is a typical nomenclature in literature to reference
this group of methods.
These are considered identical for $\Rey < 100$ \citep{Heymsfield:2010}.
We select from the literature and show in \rt{Figure}
\ref{fig:ResColumnarIceParticle} some experimental data presented by
\citet{Jayaweera:1972,Kajikawa:1973,Michaeli:1977}.

The LBM results in \rt{Figure} \ref{fig:ResColumnarIceParticle} are close to
\rt{the} \citet{Westbrook:2008} theoretical proposal for hexagonal columns in
random orientation, and below the MHKC proposals.
As expected, the LBM results for horizontal and vertical crystal orientation
lay respectively below and above the \citet{Westbrook:2008} theoretical proposal.
The difference between \rt{terminal} velocity for crystals in horizontal and
vertical orientation increase with $a_{r}$.
\rt{Since the data in Figure \ref{fig:ResColumnarIceParticle} contain results
for all the tests, some of them at different mass densities as explained, we
observe that $v_{n}$ is essentially mass independent.
Moreover, $v_{n}$ since to depend only on the aspect ratio for the analyzed
columnar ice crystals. The observed behaviours are in accordance with
\citet{Westbrook:2008} proposal.}
%

\begin{figure}
\centering
\includegraphics[scale=1.0]{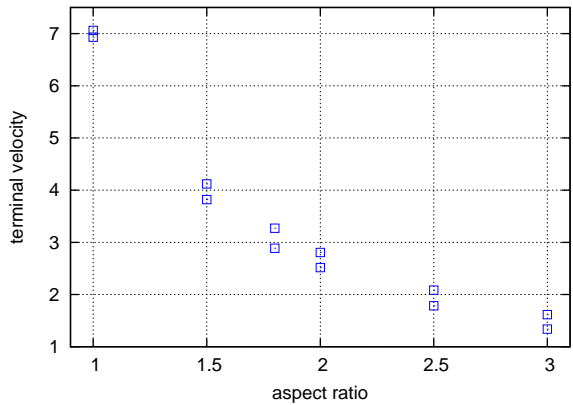}
\caption{\rt{Terminal velocity $v_{c}$ for columnar ice crystals as a function
of aspect ratio $a_{r}$. The LBM results correspond to $l=50 \mathrm{\mu m}$
with $\rho_{ice}=800\mathrm{kg \, m^{-3}}$.
Velocity is expressed in $\mathrm{cm \, s^{-1}}$, size in $\mathrm{\mu m}$.}}
\label{fig:ResVelocity_L50_rho800}
\end{figure}

\begin{figure}
\centering
\includegraphics[scale=1.0]{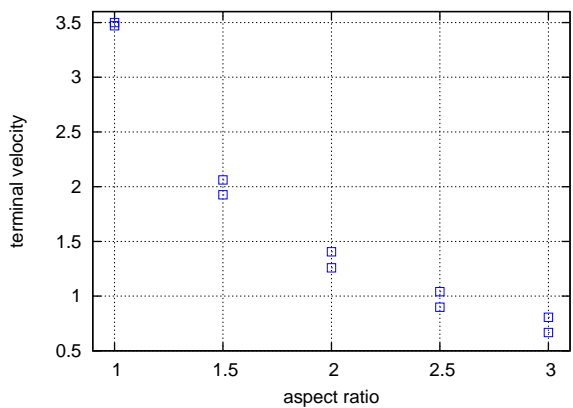}
\caption{\rt{Terminal velocity $v_{c}$ for columnar ice crystals as a function
of aspect ratio $a_{r}$. The LBM results correspond to $l=50 \mathrm{\mu m}$
with $\rho_{ice}=400\mathrm{kg \, m^{-3}}$.
Velocity is expressed in $\mathrm{cm \, s^{-1}}$, size in $\mathrm{\mu m}$.}}
\label{fig:ResVelocity_L50_rho400}
\end{figure}

\rt{In Figures \ref{fig:ResVelocity_L50_rho800} and
\ref{fig:ResVelocity_L50_rho400} we show, as it is usually presented
in the literature, the \rt{terminal} velocity for an ice crystal with
$l=50 \mathrm{\mu m}$ in the $a_{r}=\left[1,3\right]$ aspect ratios range.
The LBM results shown in Figures \ref{fig:ResVelocity_L50_rho800} and
\ref{fig:ResVelocity_L50_rho400} were obtained with
$\rho_{ice}=800\mathrm{kg \, m^{-3}}$ and $\rho_{ice}=400\mathrm{kg \, m^{-3}}$
respectively.
We conveniently rearrange the results to emphasize, for a particular crystal
length, the \rt{terminal} velocity variation as a function of the aspect ratio.
The adopted crystal length is chosen without any particular reason.}

\subsection{Implementation details}

The fluid dynamics is computed in a reference frame that moves downward with
constant velocity with respect to the lab frame. The frame velocity is set to be
equal to the expected particle's terminal velocity with respect to \rt{the}
earth.

\begin{figure}
\centering
\includegraphics[scale=0.4]{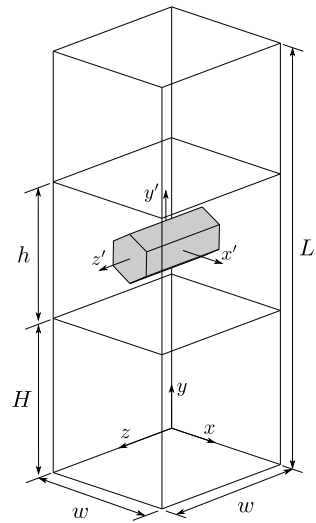}
\caption{Fluid domain scheme used in the LBM simulation. Global position of grid
refinement and rigid body.}
\label{fig:FluiDomainScheme}
\end{figure}

The computational domain, shown in \rt{Figure} \ref{fig:FluiDomainScheme}, is a
finite region of the moving frame.
At the center of the domain there is a refinement region of length $h$.
This region has an arrangement of different grid sizes. The rigid body lay
completely in the \rt{region with smallest grid size}.
Above and below the finest grid we pile grid blocks of successive increasing
grid size.
This arrangement is repeated in both sides of finest grid to reach the coarsest
grid.  We use five grid sizes along the longitudinal axis of the fluid domain as
explained in section \ref{subsec:GridRefinementMethod}.

There are at least two main reasons to analyze the problem in a constant
velocity frame.
On the one hand, being the rigid body at an approximately constant position,
relative to the computational fluid domain, reduces the computational cost in
moving the grids, since we can keep the refinement region approximately fixed in
time. On the other hand, the constant velocity frame allows us to reduce the
length of the domain.

The boundary conditions on the computational domain are, free slip on the
vertical walls, Dirichlet constant velocity on the bottom wall, and convective
boundary condition on the top wall.

The initial condition for the flow is to set an homogeneous velocity $u_{0}$
pointing upwards, meaning fluid at rest in a lab reference frame, in the whole
fluid flow domain.

For $t>0$ there is a transient flow and body movement we are not interested in.
After short time $t_s$ a stationary regime is reached.
If the velocity $u_{0}$ was chosen \rt{correctly}, the body remains roughly
static in the stationary regime.

\begin{figure}
\centering
\includegraphics[scale=1.0]{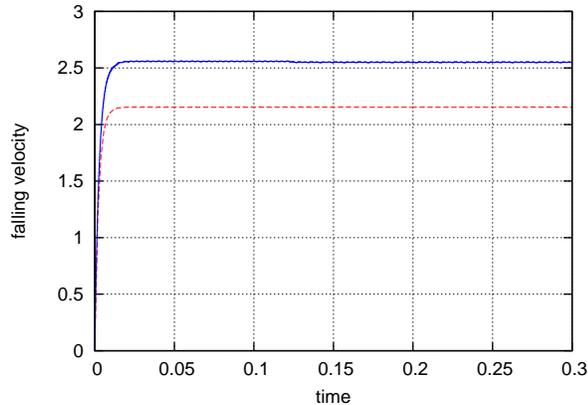}
\caption{Evolution of falling velocity for an hexagonal columns with
$l=60\mathrm{\mu m}$ and $a_{r}=2.8$. The curves show the falling velocity for
the crystal in horizontal (dashed line) and vertical sedimentation (continuum
line).
Velocity is expressed in $\mathrm{cm \,s^{-1}}$ and time in $\mathrm{s}$.}
\label{fig:ResEvolutionFallingVelocity}
\end{figure}

In Figure \ref{fig:ResEvolutionFallingVelocity} we show the evolution of
falling velocity for an hexagonal column in vertical and horizontal
sedimentation.
The results correspond to a crystal with $l=60\mathrm{\mu m}$ and $a_{r}=2.8$.
From Figure \ref{fig:ResEvolutionFallingVelocity}, the numerical results
have some minor but not negligible noise.
Then, the adopted \rt{terminal} velocity is obtained as a median velocity in
about the last $0.1\mathrm{s}$ of the simulation time.
We have made longer runs than those shown in Figure
\ref{fig:ResEvolutionFallingVelocity}, and no appreciable difference in the
terminal velocity is observed.

The rigid body is initially placed in the cross sectional center at
approximately $7w$ from the bottom wall domain. \rt{$w$ is the side length of
the computational domain, see Figure \ref{fig:FluiDomainScheme}.}
We use computational domains with relation $\frac{L}{w}\geq 12$.

The presence of near free slip walls may have a minor but not negligible
influence \rt{on} the numerical falling velocity.
For blockage ratios (defined here as $d/w$) \rt{longer} than \rt{a} certain
value, corrections should be applied to the results of numerical simulations.
We observe that for blockage ratios smaller than $5.5\%$ the influence of the
walls is negligible.
This maximum acceptable blockage value was obtained by evaluating the
interference effects on an sphere in sedimentation.
These interference effects are quantified by the relation between the
LBM obtained \rt{terminal} velocity and that obtained from theoretical
estimations in an unrestricted domain.
Differences less than $0.5\%$ between these velocities were observed for
$d/w \leq 0.055$.

The computed results shown in this paper were obtained with blockage ratios
smaller than $5.5\%$.
This configuration allow us to get $v_{n}$ by numerical evaluation of
$v_{c}$ and with $v_{r}$ obtained from Stokes equation.

\section{Conclusion and discussion}
\label{sec:ConclusionAndDiscussion}

We present in this work a Lattice-Boltzmann method to determine the dynamics of
ice crystals in the atmosphere.
Given a characterization for the shape, size and mass density of the crystal,
together with the atmospheric \rt{conditions}, it is possible to completely
determine its dynamical behavior.
The numerical method proposed provides good results for the sedimenting velocity
for the geometries, sizes and range of Reynolds number analyzed.

The LBM method takes into account the real geometry of the crystals. No
approximations, as those proposed by \citet{Bohm:1989} and \citet{Mitchell:1996}
which are widely used in the literature, are needed.
\rt{Naturally, one needs to specify the particles parameters like mass and
aspect ratio.}

For the hexagonal column crystals, the results obtained by LBM are completely
\rt{within} the dispersion region of the experimental laboratory
measurements. When
the capacitance (eq. \ref{eq:ColumnarCapacitance}, \citep{Westbrook:2008b}) is
used as \rt{a} variable, the dispersion of both experimental and LBM results
decreases noticeable. In this case a small bias of the LBM results towards the
lower end of the dispersion region can be observed.

\rt{It} was not the purpose of this paper to obtain a good fitting curve for the
terminal velocity but rather \rt{to make} a direct comparison with experimental
data.
To actually get such a curve\rt{,} more points should be computed.

By direct comparison, we see that the LBM results in Figure
\ref{fig:ResColumnarIceParticle} are close to \rt{the} \citet{Westbrook:2008}
theoretical proposal for hexagonal columns in random orientation, and below the
MHKC proposals.
As expected, the LBM results for horizontal and vertical crystals orientation
lay respectively below and above the \citet{Westbrook:2008} theoretical proposal.
Also, the difference between \rt{terminal} velocity for crystals in horizontal
and vertical orientation increase with $a_{r}$.

As the \rt{final} message of this work, we want to emphasize that a great deal
of problems of interest in relation to the physics of the atmosphere can be
analyzed via LBM methods.
As regards falling ice particles\rt{,} one could study different geometries, or
different values of parameters.
One could get statistical characterizations for cases where laboratory
experiments become very difficult. One could also use LBM simulations to test
proposed models, the sensitivity of results to the parameters involved, etc.

\section*{Acknowledgments}
The author wants to thank Nesvit E. Castellano, Rodrigo E. B\"urgesser and
Omar E. Ortiz for useful discussions and contributions.
N. E. Castellano and R. E. B\"urgesser brought the author's attention to this
subject.
J. P. Giovacchini is a fellowship holder of CONICET (Argentina).  This work was
supported in part by grants 05-B454 of SECyT, UNC and PIDDEF 35/12 (Ministry of
Defense, Argentina).

\appendix

\section{Grid convergence verification for $a_{r}=2$ columnar ice crystal}
\label{AppA:GridConvergence}

Not to increment the computational cost more than necessary for the purposes of
this work, we study in this appendix how the grid parameter $\delta x$ affects
the outcome of our simulations.
More precisely, we want to determine an acceptable value of the parameter
$d  /\delta x$, the amount of lattice nodes falling inside the crystal diameter,
so that no substantial change in the outcome occurs if this parameter is
increased.

\begin{figure}
\centering
\includegraphics[scale=1.0]{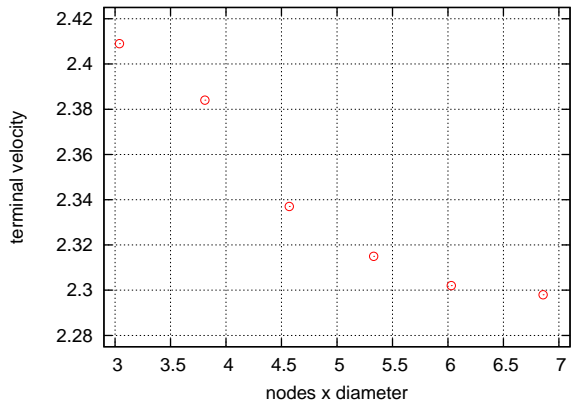}
\caption{\rt{Terminal} velocity $v_{c}$ as a function of nodes per cylinder
diameter.
Circular cylinder with $a_{r}=2$ and $l=50\,\mathrm{\mu m}$ in horizontal
sedimentation.}
\label{fig:ResConvergenceVelCilinders}
\end{figure}

As a typical example we analyze the case of a cylinder sedimenting with
horizontal orientation with parameters $a_r = 2,$ $l = 50 \mathrm{\mu m}$ and
$\rho_{ice} = 650 \mathrm{kg\,m^{-3}}$.
Figure \ref{fig:ResConvergenceVelCilinders} shows the terminal sedimentation
velocity obtained for different discretizations.
In all simulations done in this work we adopt a discretization range
$5.5 < \frac{2a}{\delta x} < 6.1$ which we consider an acceptable
discretization.

We choose a cylinder for this test because there is experimental data to compare
with and because the geometry is similar to that of hexagonal cross section
columns.
Experimental results for the sedimenting velocity for horizontal cylinders can
be found in \cite{Westbrook:2008} in non dimensional form.
We compare these results with the ones we obtain with LBM.
From Westbrook (2008), $v_{c}\simeq 2.3 \mathrm{cm\,s^{-1}}$
($v_{n} \approx 1.3$),
for the horizontal sedimentation and $v_{c}\simeq 2.65 \mathrm{cm\,s^{-1}}$
($v_{n} \approx 1.5$) for the vertical sedimentation.
With LBM and $\frac{2a}{\delta x} \approx 5.5$ we obtain
$v_{c} = 2.64 \mathrm{cm\,s^{-1}}$ for vertical and
$v_{c} = 2.31 \mathrm{cm\,s^{-1}}$ for horizontal case.

\section{LBM results}
\label{AppB:LBMresults}

In this appendix we present the LBM results showed in section
\ref{subsec:ResAndDisc}.
In tables \ref{AppB:Res_ar_In_1-2} and \ref{AppB:Res_ar_In_2-3} we present the
results for $a_{r}\in[1,2]$ and $a_{r}\in(2,3]$ respectively.
The selected $l$ and $a_{r}$ values are in the range of data measured by
\citet{Burgesser:2016}.

\begin{table}
\centering
 \begin{tabular}{ccccccc}
  \hline
  Test & $a_{r}$ & $l$ & $v_{c_{v}}$ & $v_{c_{h}}$ & $v_{r}$ & $C$ \\
\hline
1 & $2$ & $50$ & $2.30$ & $2.02$ & $1.465$ & $18.83$ \\
\hline
2 & $1.5$ & $37.5$ & $2.49$ & $2.28$ & $1.782$ & $16.58$ \\
\hline
3 & $1.2$ & $20$ & $0.97$ & $0.90$ & $0.731$ & $10.09$ \\
\hline
4 & $1.75$ & $40$ & $2.35$ & $2.06$ & $1.555$ & $16.2$ \\
\hline
5 & $1.0$ & $23$ & $1.51$ & $1.45$ & $1.207$ & $13.0$ \\
\hline
6 & $1.1$ & $17$ & $0.73$ & $0.69$ & $0.570$ & $9.05$ \\
\hline
7 & $1.9$ & $35$ & $1.49$ & $1.32$ & $0.968$ & $13.55$ \\
\hline
8 & $1.4$ & $30$ & $1.64$ & $1.53$ & $1.205$ & $13.81$ \\
\hline
9 & $1.8$ & $50$ & $3.27$ & $2.88$ & $2.176$ & $19.94$ \\
\hline
10 & $1.7$ & $47$ & $3.18$ & $2.80$ & $2.125$ & $19.357$ \\
\hline
11 & $1.15$ & $25$ & $1.50$ & $1.39$ & $1.150$ & $12.95$ \\
\hline
12 & $1.25$ & $32$ & $2.24$ & $2.04$ & $1.651$ & $15.76$ \\
\hline
13 & $1.6$ & $42$ & $2.72$ & $2.46$ & $1.885$ & $17.9$ \\
\hline
14 & $1.0$ & $15$ & $0.642$ & $0.637$ & $0.513$ & $8.48$ \\
\hline
15 & $1.3$ & $28$ & $1.57$ & $1.49$ & $1.186$ & $13.47$ \\
\hline
16 & $1.5$ & $45$ & $3.36$ & $3.09$ & $2.416$ & $19.90$ \\
\hline
17 & $1.0$ & $50$ & $7.06$ & $6.93$ & $5.704$ & $28.27$ \\
\hline
18 & $1.5$ & $50$ & $4.12$ & $3.82$ & $2.983$ & $22.11$ \\
\hline
19 & $2.0$ & $50$ & $2.52$ & $2.81$ & $1.803$ & $18.83$ \\
\hline
20 & $1.0$ & $50$ & $3.5$ & $3.47$ & $2.85$ & $28.27$ \\
\hline
21 & $1.5$ & $50$ & $2.06$ & $1.92$ & $1.491$ & $22.11$ \\
\hline
22 & $2.0$ & $50$ & $1.40$ & $1.26$ & $0.902$ & $18.83$ \\
\hline
23 & $1.25$ & $30$ & $0.954$ & $0.893$ & $0.725$ & $14.77$ \\
\hline
24 & $1.75$ & $30$ & $0.610$ & $0.555$ & $0.411$ & $12.15$ \\
\hline
\end{tabular}
\caption{LBM results for $a_{r}\in[1,2]$.
$l$ and $C$ are in $\mathrm{\mu m}$, velocity in $\mathrm{cm\,s^{-1}}$.
\rt{$v_{c_{v}}$ and $v_{c_{h}}$ are the \rt{terminal} velocities for crystals in
vertical and horizontal sedimentation respectively.
All the results were obtained with $\rho_{ice}=800\mathrm{kg\,m^{-3}}$, except
the tests labeled as Test 1 and 20-24 were obtained with
$\rho_{ice}=650\mathrm{kg\,m^{-3}}$ and $\rho_{ice}=400\mathrm{kg\,m^{-3}}$
respectively}}
\label{AppB:Res_ar_In_1-2}
\end{table}

\rt{The results labeled as Test 1 and 20-24 in table \ref{AppB:Res_ar_In_1-2}
were obtained with $\rho_{ice}=650\mathrm{kg\,m^{-3}}$ and
$\rho_{ice}=400\mathrm{kg\,m^{-3}}$ respectively.
The results labeled as Test 20, 21, and 22 are shown exclusively in the Figures
\ref{fig:ResVelocity_L50_rho400} and \ref{fig:ResColumnarIceParticle}.
The results in Test 1, 23, and 24 are only shown in the Figure
\ref{fig:ResColumnarIceParticle} where the results are expected to be mass
independent.}

\begin{table}
\centering
\begin{tabular}{ccccccc}
\hline
$Test$ & $a_{r}$ & $l$ & $v_{c_{h}}$ & $v_{c_{v}}$ & $v_{r}$ & $C$ \\
\hline
1 & $2.05$ & $27$ & $0.69$ & $0.80$ & $0.503$ & $10.036$ \\
\hline
2 & $2.1$ & $40$ & $1.45$ & $1.72$ & $1.057$ & $14.67$ \\
\hline
3 & $2.2$ & $35$ & $1.06$ & $1.22$ & $0.743$ & $11.45$ \\
\hline
4 & $2.5$ & $52$ & $1.92$ & $2.25$ & $1.29$ & $17.42$ \\
\hline
5 & $2.8$ & $60$ & $2.15$ & $2.56$ & $1.395$ & $18.99$ \\
\hline
6 & $2.3$ & $55$ & $2.44$ & $2.83$ & $1.692$ & $19.23$ \\
\hline
7 & $2.025$ & $31$ & $0.95$ & $1.07$ & $0.678$ & $11.59$ \\
\hline
8 & $2.15$ & $33$ & $0.97$ & $1.12$ & $0.689$ & $11.95$ \\
\hline
9 & $2.35$ & $48$ & $1.79$ & $2.12$ & $1.238$ & $16.604$ \\
\hline
10 & $2.6$ & $57$ & $2.14$ & $2.57$ & $1.447$ & $18.724$ \\
\hline
11 & $2.4$ & $46$ & $1.59$ & $1.87$ & $1.094$ & $15.74$ \\
\hline
12 & $2.7$ & $65$ & $2.61$ & $3.17$  & $1.753$ & $20.95$ \\
\hline
13 & $3.0$ & $80$ & $3.41$ & $4.07$ & $2.176$ & $24.48$ \\
\hline
14 & $2.25$ & $62$ & $3.19$ & $3.71$ & $2.239$ & $21.93$ \\
\hline
15 & $2.9$ & $70$ & $2.76$ & $3.31$ & $1.777$ & $21.77$ \\
\hline
16 & $2.5$ & $50$ & $1.78$ & $2.08$ & $1.198$ & $16.75$ \\
\hline
17 & $3.0$ & $50$ & $1.34$ & $1.61$ & $0.850$ & $15.3$ \\
\hline
18 & $3.0$ & $50$ & $0.667$ & $0.806$ & $0.425$ & $15.3$ \\
\hline
19 & $2.5$ & $50$ & $0.898$ & $1.04$ & $0.599$ & $16.75$ \\
\hline
20 & $2.25$ & $30$ & $0.378$ & $0.436$ & $0.262$ & $10.61$ \\
\hline
21 & $2.75$ & $30$ & $0.275$ & $0.328$ & $0.180$ & $9.58$ \\
\hline
\end{tabular}
\caption{LBM results for $a_{r}\in(2,3]$.
$l$ and $C$ are in $\mathrm{\mu m}$, velocity in $\mathrm{cm\,s^{-1}}$.
\rt{$v_{c_{v}}$ and $v_{c_{h}}$ are the \rt{terminal} velocities for crystals in
vertical and horizontal sedimentation respectively.
All the results were obtained with $\rho_{ice}=800\mathrm{kg\,m^{-3}}$, except
the tests labeled as Test 18-21 that were obtained with
$\rho_{ice}=400\mathrm{kg\,m^{-3}}$.}}
\label{AppB:Res_ar_In_2-3}
\end{table}

\rt{The results labeled as Test 18-21 in table \ref{AppB:Res_ar_In_2-3}
were obtained with $\rho_{ice}=400\mathrm{kg\,m^{-3}}$.
The results labeled as Test 18 and 19 are shown exclusively in the Figures
\ref{fig:ResVelocity_L50_rho400} and \ref{fig:ResColumnarIceParticle}.
The results in Test 20 and 21 are only shown in the Figure
\ref{fig:ResColumnarIceParticle}.}

\bibliographystyle{wileyqj}

\end{document}